\documentclass[iop]{emulateapj}
\usepackage{graphicx}
\usepackage{epstopdf}
\usepackage{natbib}
\usepackage{amsmath}

\usepackage{url}

\submitted{Submitted: 2013 April 1}
\journalinfo{{\sc The Journal of Speculative Astronomy} }

\shortauthors{Davenport}
\shorttitle{UFOs in the Time Domain}

\begin{document}

\title{Unidentified Moving Objects in Next Generation Time Domain Surveys}
\author{James R. A. Davenport\altaffilmark{1}}
\altaffiltext{1}{Department of Astronomy, University of Washington}

\begin{abstract}
Existing and future wide-field photometric surveys will produce a time-lapse movie of the sky that will revolutionize our census of variable and moving astronomical and atmospheric phenomena. As with any revolution in scientific measurement capability, this new species of data will also present us with results that are sure to surprise and confound our understanding of the cosmos. While we cannot predict the unknown yields of such endeavors, it is a beneficial exercise to explore certain parameter spaces using reasonable assumptions for rates and observability. To this end I present a simple parameterized model of the detectability of unidentified flying objects (UFOs) with the Large Synoptic Survey Telescope (LSST). I also demonstrate that the LSST is well suited to place the first systematic constraints on the rate of UFO and extraterrestrial visits to our world.
\end{abstract}

\section{Introduction}

The study of unidentified flying objects (UFOs) and extraterrestrials (ETs) is frequently the purview of non-academic circles. This is the result of both academic taboos, and the great difficulty in conducting rigorous controls when gathering data. 
It is worth noting, however, that astronomers have expended significant effort in searching for and understanding ETs in the past century. In 1974, a simple message was broadcast from Arecibo Observatory towards the globular cluster M13 \citep{1975Icar...26..462}. More recently the SETI@home project \citep{setiathome} has utilized more than 6 million volunteer computers to search over 160 TB of Arecibo data for signs of ET radio broadcasts. 

While most searches for observable indicators of ETs have used radio telescopes, a notable counter-example is the SEVENDIP program designed to search for very short timescale pulses of optical light from nearby FGKM stars \citep{2001SPIE.4273..104W}. 
Studies constraining the so-called Drake Equation, which parameterizes the probability of life,  continue to be pursued \citep[e.g.][]{2004AsBio...4..225C,2008ASPC..395..213D}. Perhaps most encouraging to the long term study of ETs, Astrobiology has become a rich field of study, and today  supports a very well regarded peer-reviewed journal by the same name.

The search for UFOs enjoys less legitimacy in the astronomical community, however. Most  accounts of UFO observations have been the result of by-eye identification, usually by the general public. Many such incidents are later correlated with atmospheric phenomena, military or civilian aircraft flights, or any number of other mundane terrestrial activities. A culture of paranoia, myth, astrology, ridicule, and obsession has subsequently grown up around the study of UFOs. This provides a significant, though understandable, barrier towards their study and discourse within the academic sphere.

We gladly accept the study of astrobiology as fruitful and beneficial, that indications of life may someday soon be discovered outside of our planet or even our solar system. So too must we accept that it is indeed possible, regardless of our own discomfort or preconceived notions on the subject, that ETs could exist and be visiting our world. While the probability may be quite small, it is decidedly non-zero.

As the era of ``big data'' and ``time-domain'' astrophysics blossoms, we should be ever mindful of the ancillary or atypical questions that can be answered with these new datasets. The next generation of wide-field astronomical surveys will provide an unprecedented opportunity to monitor the sky and systematically search for signs of ET or UFO activity. In this short April Fools letter I highlight the Large Synoptic Survey Telescope, as well as many smaller surveys, as being able to place the first quantifiable upper limits on the rates of such phenomena.

\begin{figure*}[t]
\centering
\includegraphics[width=3in]{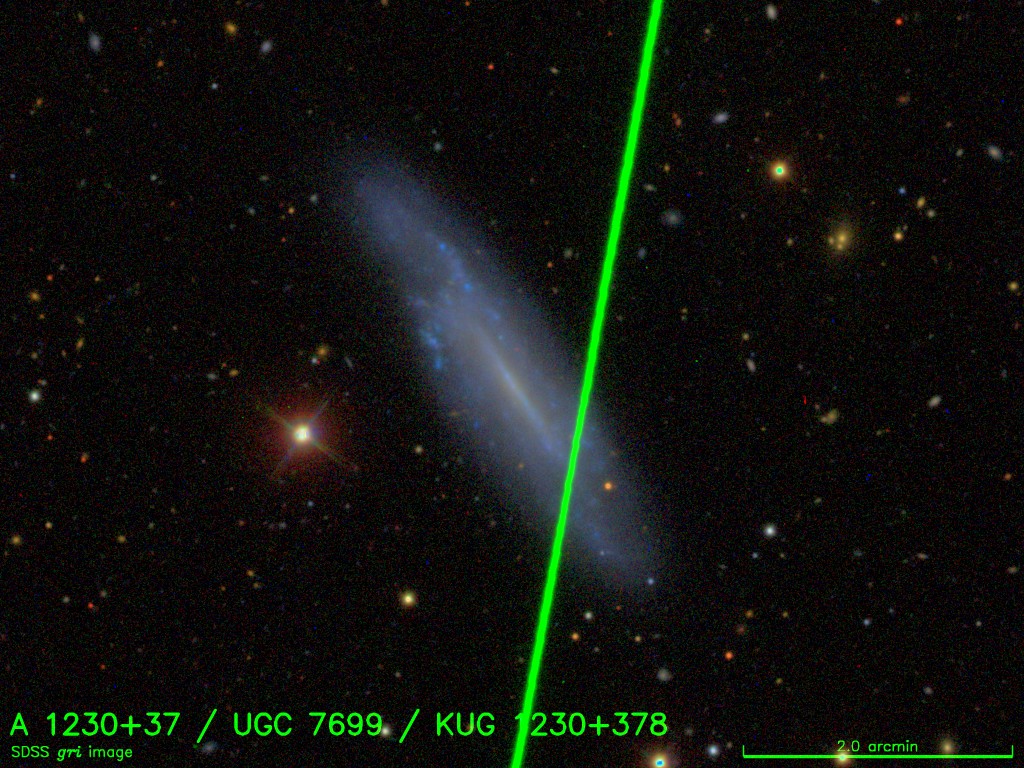}
\includegraphics[width=3in]{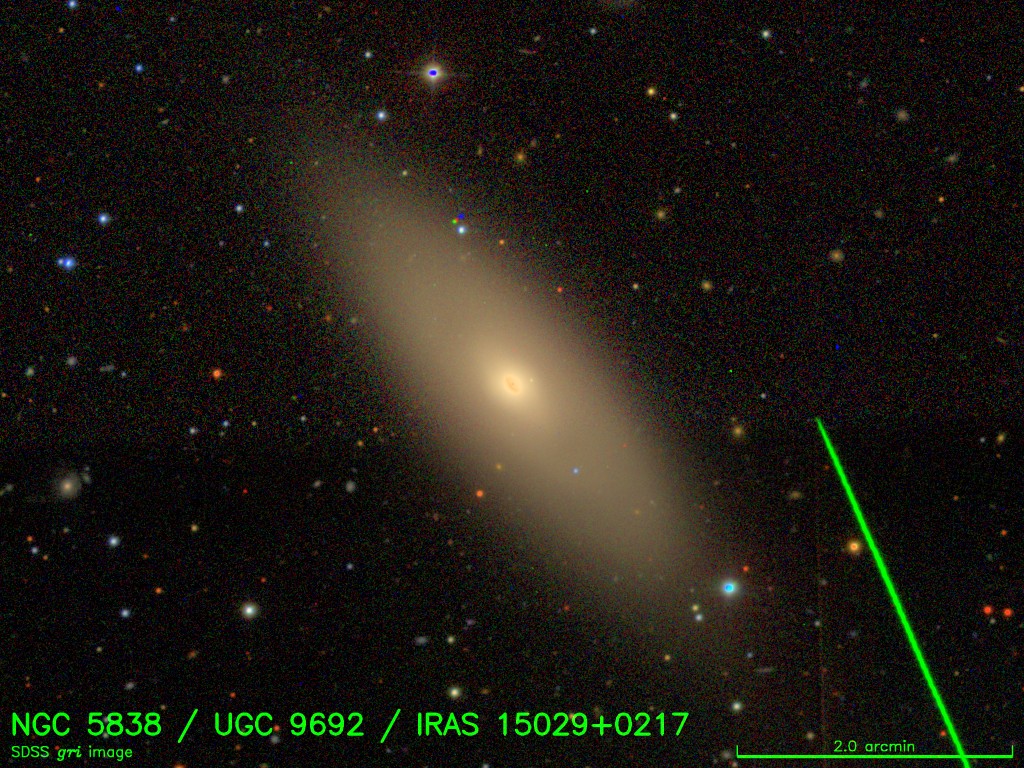}
\caption{Two  meteorites from the SDSS DR6, each observed in a single 52 second image, shown as green ``streaks'' or ``trails''. Images copyrighted 2006 David W. Hogg, Michael R. Blanton, and the Sloan Digital Sky Survey Collaboration}
\label{asteroids}
\end{figure*}


\section{Constraints from Photometric Surveys}
The Large Synoptic Survey Telescope \cite[LSST;][]{lsst} is an revolutionary project to create a ``movie of the sky''. By mapping the full night sky every three days, LSST will accumulate $\sim$100 frames spread across six photometric bands ($ugrizy$) over the course of ten years. A freeze-frame strategy will be employed, taking two 15 second exposures at each pointing to facilitate moving object detection and characterization. The large 6.7-meter (effective) aperture of the LSST will provide single-visit imaging to a depth of $r\sim24.5$ over a 9.6 deg$^2$ field of view. 

Such a machine could search for UFOs in two regimes: above and within the atmosphere. Traditional low-altitude encounters with UFOs, commonly described as ``sightings'' by members of the public, would undoubtedly be so bright as to saturate the LSST camera in any filter. Orbiting or passing UFOs that reflect even a small amount of sunlight could also be strongly detected. These detections would be more akin to observations of meteorites, asteroids, or Earth-orbiting satellites, appearing as a bright streak across the field of view. In this section I describe some of the guidance we can gain from examining previous large scale photometric surveys for unusual objects, as well as the limited applicable data on UFO sightings. 

\vspace{0.25in}

\subsection{Lessons from Existing Surveys}
The increasing number of satellites in Earth orbit has created a kind of light pollution for astronomical surveys \citep{1992Msngr..67...53F}. This additional noise source must be accounted for, both for the characterization of traditional survey goals, and to identify true outliers. To sensibly predict what the signatures of observing an UFO with the LSST might look like, we can turn to the available imaging from existing surveys such as the Sloan Digital Sky Survey \citep[SDSS;][]{york2000}. The primary data goal of the SDSS was to produce a single-epoch image for a quarter of the sky, with nearly simultaneous imaging in five photometric passbands. A continuous scanning technique was used, letting stars drift linearly across the field of view. This produced $\sim$52 second exposures in each band in the order $r,i,u,z,g$. The small time delay between images produced value-added information for moving objects such as asteroids that can be used to measure instantaneous orbital trajectories. 

While the survey operation and cadence of LSST will be vastly different from that of SDSS, the analysis and software requirements are in many ways similar. Object detection, measurement, and identification from repeated observations must be handled systematically and without human intervention. The result is a myriad of secondary data products, flags, characteristics, and measurements that would complicate the data to the point of uselessness without adequate prescriptions, as was provided in the SDSS documentation.

The SDSS observed many chance\footnote{if indeed there is such a thing as chance} events throughout its operation, which are commonly filtered out using flag and quality cuts appropriate for normal science operations. These transient objects are known to have included meteorites, airplanes, asteroids, and satellites. 

Two examples of suspected meteorites that were observed by the SDSS during normal imaging are shown in Figure \ref{asteroids}. The importance of proper identification and removal of such events is clear. The left panel of Figure \ref{asteroids} shows a meteorite that crossed the entire CCD field of view during the $r$-band exposure, obscuring several stars, and contaminating the resolved galaxy UGC 7699. The right panel shows a similar event, with a meteorite partially obscuring the frame of UGC 9692. 

\begin{figure*}[!ht]
\centering
\includegraphics[width=2.5in]{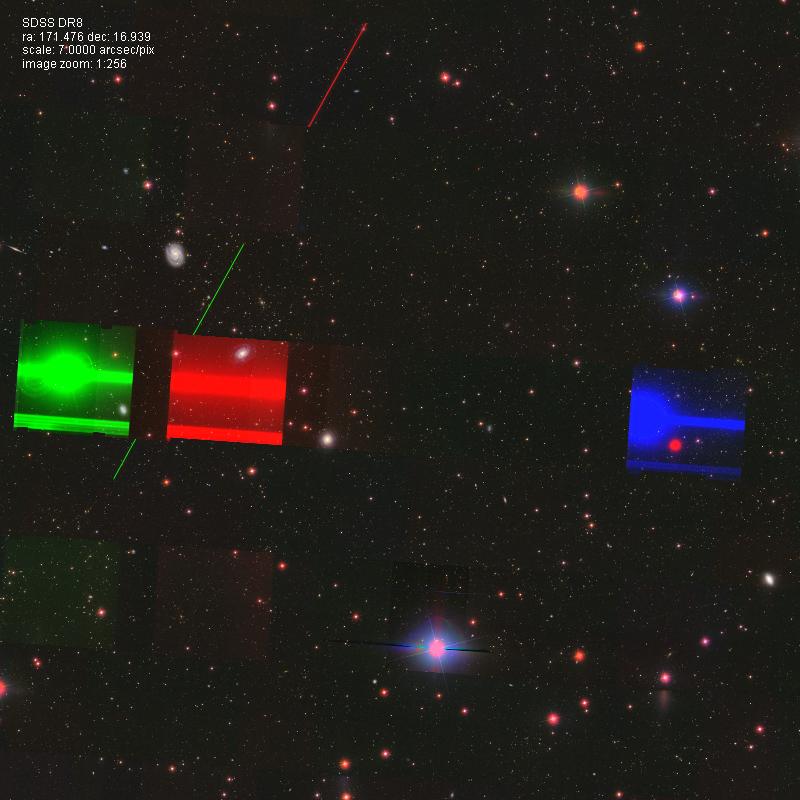} \hspace{0.1in}
\includegraphics[width=3.35in]{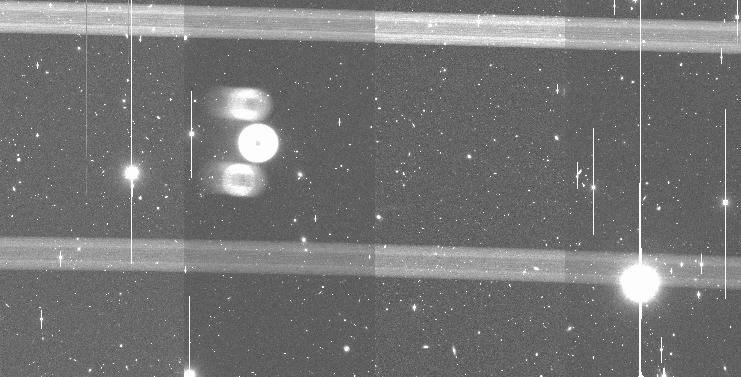}
\caption{Left: A dramatic image artifact in the SDSS DR6, possibly an airplane, noted by users of the GalaxyZoo program. Right: A verified airplane imaged by the MACHO project.}
\label{plane}
\end{figure*}

In the left panel of Figure \ref{plane}, a highly saturated panchromatic image artifact was seen across several degrees. This event, attributed to an airplane passing into the field of view, was identified by public users of the GalaxyZoo SDSS database \citep{galaxyzoo}. Excess background flux is also present in neighboring fields within the image, due to scattered light from the airplane. Contiguous imaging for the SDSS required two passes over each field to fill the gaps between the CCDs. As a result, the excess flux from the airplane is seen on every other horizontal row in the image. The narrow green ($r$-band) and red ($i$-band) streak, probably an asteroid, was not causally associated with the airplane. 

Smaller aperture time domain surveys are also a bountiful resource for identifying moving or transient objects of interest. These provide shallower depth imaging over a very wide area, and would be better suited for in-atmosphere UFO observations. For example, the right panel of Figure \ref{plane} shows a verified airplane captured in the MACHO imaging using the 1.27-m MSO telescope \citep{macho}. An (incomplete) listing of notable existing small aperture surveys also includes OGLE \citep{ogle}, ASAS \citep{asas}, Catalina \citep{catalina}, PTF \citep{ptf}, and LINEAR \citep{linear}.  

These dramatic examples belong to of a large family of known Earth-bound contaminants in photometric surveys. By understanding their observed properties (e.g. color, duration, apparent motion, luminosity throughout the event, frequency across the sky) we may begin to automate the accurate identification and removal of each type event. This will enable us to identify truly unique or unexpected events, such as ET visits and UFO sightings.

A systematic search through these existing databases would also provide useful constraints on brighter UFO events, as well as a robust training set for moving object characterization. Most contaminating streaks seen in imaging surveys appear linear across the field of view, as in Figures \ref{asteroids} and \ref{plane}. However, piloted UFOs may be best detected by their non-linear trajectories in images.


\subsection{UFO Reference Data}

In 1992, a dramatic UFO was widely seen over Chile, described as a slow moving and very bright object in the sky. Notably, photographs were even published by professional astronomers \citep{1992Msngr..67...56H}. This event was quickly suggested to be the result of debris burning up from a recent spacecraft launch \citep{1992Msngr..68...42B}, and many similar events have been known to occur in the skies over Chile. The true origin was never verified. Coincidence?

\begin{figure}[]
\centering
\includegraphics[width=3.45in]{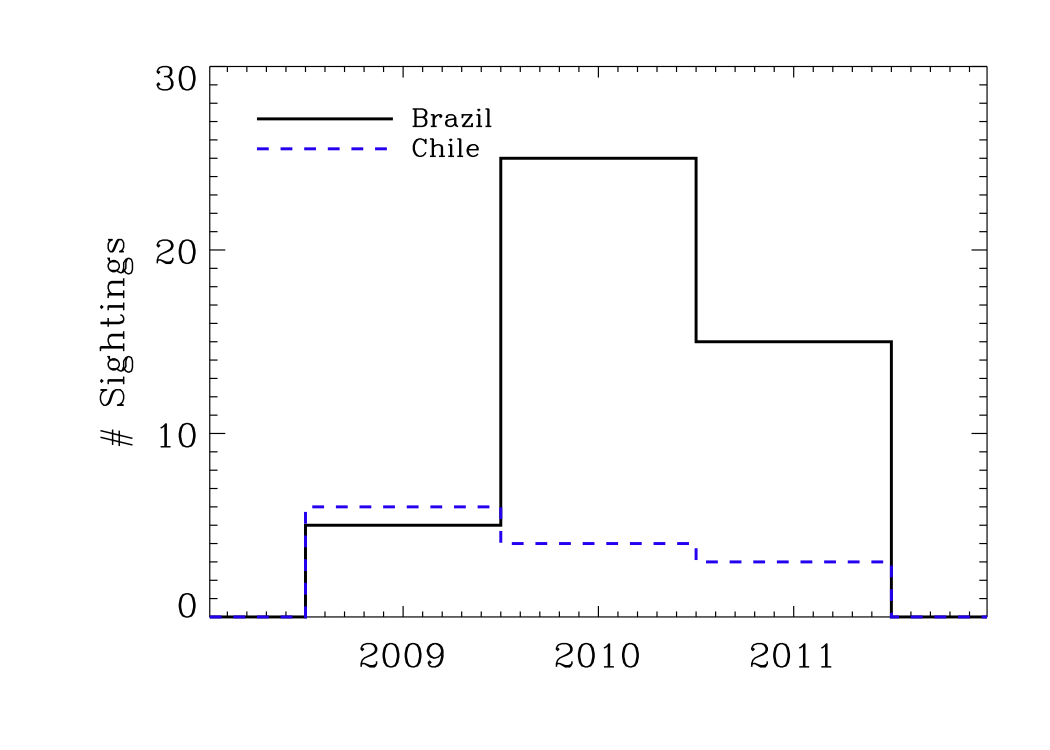}
\caption{Numbers of reported UFO sightings for Chile and Brazil from 2009 to 2011.}
\label{years}
\end{figure}

Since no official IAU catalog or repository for such events exists, UFO researchers are forced to either laboriously collect data (often in the form of testimonials many years after the fact) or rely on the publication of UFO/ET sightings in local media, books, or police reports. Some efforts have been made to collect large samples of testimonials and reports to form databases that are searchable by date, location, event type, etc. One such resource is the Mutual UFO Network (MUFON) online database\footnote{\url{http://www.mufon.com/}} that contains over 30,000 indexed sightings worldwide. Figure \ref{years} shows the total numbers of reported UFO sightings for two South American countries, Brazil and Chile, between 2009 and 2011 using the MUFON database. While the long-term trends are not clear over so short a time span, one notable result is that the gross rate {\it per capita} of UFO sightings is three times higher in Chile than in Brazil. I also note that  these numbers likely represent lower-limits on the actual UFO/ET visitation rate in these countries, as they are all the result of naked-eye sightings, and have no constraint on orbiting spacecraft. Since the US run MUFON database service may also not be  well publicized in non-English speaking countries, the efficiency of collecting all UFO reports is also likely to be low. 

It is also worth mentioning that rates of UFO and ET encounters have been shown to vary based on geographic location. \citet{johnsonufo} demonstrated a heightened rate of UFO sightings (including some alarmingly detailed encounters) near US nuclear facilities. This indicates that the UFO phenomena may have an intelligent origin, and that their visitation may be related to our technological advancement. Whether these UFOs are of ET or terrestrial origin remains unknown. It is also unclear how ETs would affect their flight paths and visitation rates given the knowledge that deeper sky monitoring facilities such as LSST were present at a certain location. In other words, would ETs intentionally avoid certain orbital approaches, flight paths, or landing in specific areas in order to minimize detection by humankind? Being able to measure a change in visitation behavior would give us the first intriguing glimpse of extraterrestrial psychology.

\section{Simple Model}
Due to the paucity of detailed scientific investigations on the nature of UFO sightings, we essentially lack any quantitative data to explore parameter space. Naturally you are encouraged, dear reader, to get back to work before someone asks you {\it what on Earth} you are reading! The author will however trudge on, hope of fame and recognition thus resigned.

In this section I present a simple toy model of UFO detections to demonstrate some of the parameters that must be considered. While we do not have any publicly available information of the nature of UFO vehicles or their specifications, it is sensible to parameterize UFO sightings by observable quantities. This is conceptually identical to characterizing the observable parameter space for astrophysical transients \citep[e.g.][]{rau2009}. 

\begin{figure}[]
\includegraphics[width=3.45in]{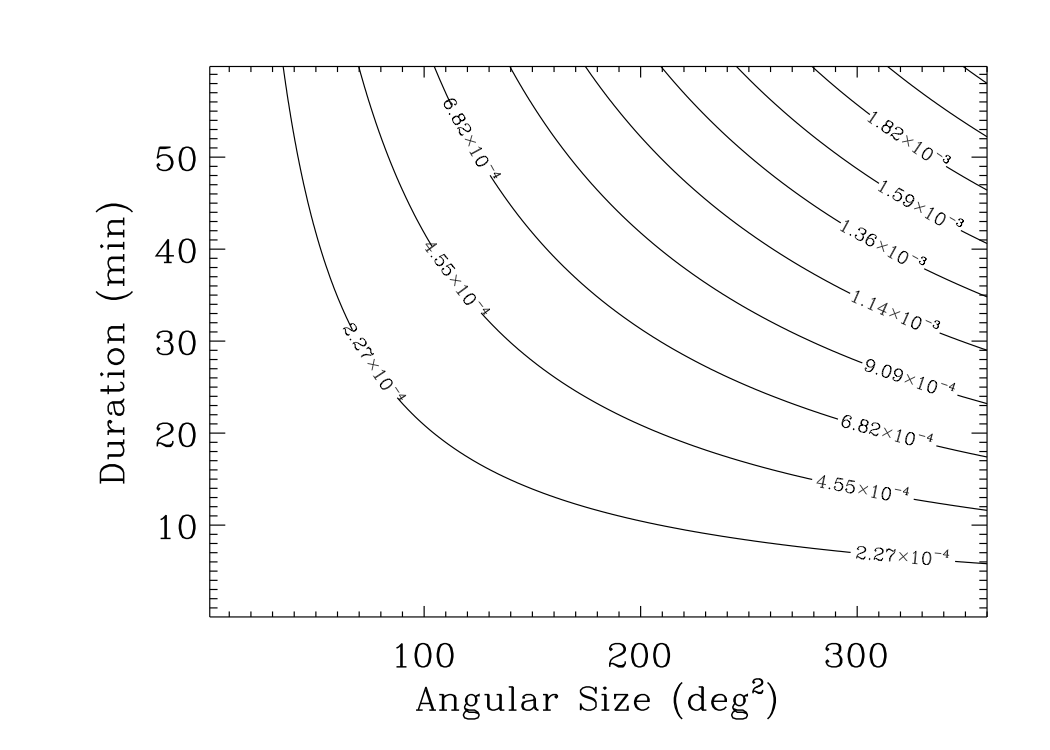}
\caption{Probability surface for detecting a single UFO  in 10 years of LSST, assuming a total of 200,000 LSST frames and 50 UFOs, as a function of the UFO angular size and duration. The maximum detection probability is 0.23\%.}
\label{prob}
\end{figure}

Consider a distribution function of UFO events, $f(\Omega,\tau,L,\vec{x},\dot{\vec{x}})$ that is a function of their apparent size $\Omega$, event duration $\tau$, brightness $L$, position $\vec{x}$, and trajectory $\dot{\vec{x}}$. The probability for detecting a UFO in the LSST as a function of event duration and event size on the sky can be described with the following generalized equation:
\begin{equation}
P_U \propto \left(\frac{\Omega_{L} \Omega_{U}}{(4\pi)^2}\right) \left( \frac{\tau_L \tau_U}{t_{day}^2}\right)N_L N_U\,,
\end{equation}
where the total number of LSST images is $N_L = 200,000$, the LSST exposure time is $\tau_L = 15$ seconds, the LSST field of view is $\Omega_L = 9.6$ deg$^2$ in sterradians , and the assumed total number of UFOs overhead in the course of the LSST mission is $N_U = 50$. As the angular size $\Omega_U$ or duration $\tau_U$ of the UFO event increases, so too does the probability of detecting the event. This general probability manifold is shown in Figure \ref{prob}.

A conservative estimate (aka total shot in the dark guess) on the number of UFO events $N_U=50$ has been used for the entire 10 years of the LSST mission. The maximum detection probability for long duration and reasonably large events in this model is $0.23\%$. The detection probability thus reaches 100\% if the total number of bright observable UFOs/ETs in 10 years is on order of 20,000. This model only describes our known unknowns. Accordingly, we have no way of knowing if we now know all the known unknowns or not. However, I invite the community to assist with a fully Bayesian approach to this problem.

\section{Obfuscation}

Once LSST has completed 10 years of observations, it may be that no UFO candidates can be found in the dataset. Analysis of existing surveys may also yield a null-result. This of course can either be due to a true lack (or very low occurrence rate) of ET visitors to our world, or from the surreptitious efforts of an unseen oligarchy preventing their discovery. 

Several major projects already coordinate their operations with various government agencies. The APOLLO project \citep{apollo}, as with most laser-guided adaptive optics facilities, must ensure they do not shine powerful lasers as aircraft. Pan-STARRS, in many ways a precursor to LSST, has had to endure military embargoes on the survey footprint for some years now. In a striking example from the world of large datasets, Google has been inundated with requests for removal or censorship of information from a host of governments.\footnote{\url{http://www.google.com/transparencyreport/}}

A satellite and space debris tracking database is maintained by the US Government, overseen by the United States Strategic Command (USSTRATCOM)\footnote{\url{https://www.space-track.org}}. Limited access to this database is granted to the public, and may be revoked or censored at any time. We therefore currently have no way to ensure a complete census of ``known'' satellites that might contaminate our study of UFOs/ETs.

It is thus conceivable that government organizations could place similar restrictions on the flow of photometric information from the LSST. There is no guarantee that this behavior will be made public, especially if the justification for the censorship is related to ETs.

\section{Conclusions}
Dawn has risen on a new era of astronomy, whereby statistics and large datasets are among our most potent tools for probing the true nature of the universe. We are called to rise up by the rooster of knowledge, crowing at the dawn, to awaken and learn as much from our wealth of data as possible. If UFOs exist and are of ET origin, then this dawn may shed early light on their properties, providing the first estimates of the visitation behavior of extraterrestrial visitors to our pale blue dot.

While projects such as the LSST may bring the survey data era to UFO studies, care must be taken to account for human contamination in the detection of such events. Moving objects seen in LSST will need to be checked against known solar system objects from the Minor Planet Center \citep{mpc} and known satellites and space debris from USSTRATCOM. 

Statistical trends of UFO or ET activity may be seen not by scientists, but by the public. Projects like GalaxyZoo, when applied to LSST data, may reveal some of the most compelling evidence for ET measurements. One imagines the difficulty a software pipeline designed for stationary objects or linear motion might have in cataloging the detected motions from a passing spacecraft under intelligent control. Human eyes would pick out such trajectories quickly from single exposures.

Intriguingly, there exists opportunities to combine the suggestive data from LSST with time honored methods of UFO studies, such as eyewitness testimonials. By correlating the mentions of UFO ``close encounters'' \citep{hynek1972} in South America from online social media services, such as Twitter, Facebook, or YouTube, with the nightly reports of transient or sporadic non-photometric events in LSST, we may be able to create a robust detection algorithm for UFOs. Conversely, we could use LSST to {\it disprove} hucksters and fanatics.

I note that exotic means of preventing such discoveries may be employed, by either human or ET influence. Apropos, the massive LSST data stream will provide an ideal hiding place for complex codes or secret signals. One trivial example comes quickly to mind: A nefarious agency could in principle place a simple spotlight in low Earth orbit, perhaps deriving its illuminating power from simply reflected sunlight. Such a platform would be useable as a signal lamp, beaming Morse code messages that would appear as ``dots'' and ``dashes'' across the LSST frames. 

Similarly, ETs may have very sophisticated means of avoiding detection from the LSST, such as cloaking devices. I have also pointed out for the first time that the discovery of a change in the ET visitation rate due to the threat of discovery by the LSST would be the first study of ET psychology. This would also provide clear evidence that these aliens are not simply watching us and stealing the occasional livestock, but instead are intimately aware of and connected to our daily affairs. Perhaps they will even read this short manuscript, and I heartily welcome any ET commentary or feedback on this work, provided it is transmitted in a language that mere humans may comprehend.

The promise of being able to constrain the ET visitation rate should not be ignored. Ultimately we cannot account for every possible scenario or cause, but there is clear need to simulate such events while developing software for the LSST. It would be a benefit to researchers of many persuasions if the LSST pipeline would provide real-time catalogs of ``smears'', ``tails'', or unexpected image saturations. 

\acknowledgements
I would like to convey my appreciation to FOIL {\sc co}, makers of high quality aluminum headwear since June 1947. Opinions and sentiments expressed herein have been intended in jest, and in no way represent any institution or governmental agency.

\end{document}